\documentclass[useAMS,usenatbib]{mn2e}
\usepackage{times}

\usepackage{amssymb}
\usepackage{./aas_macros}
\usepackage{graphicx}
\graphicspath{{images/}}
\usepackage{color}
\usepackage{natbib}

\begin{document}

\title{Modulating the magnetosphere of magnetars by internal magneto-elastic
oscillations}
\author[Michael Gabler, Pablo Cerd\'a-Dur\'an, Nikolaos Stergioulas, Jos\'e
A.~Font and Ewald M\"uller]
{Michael Gabler$^{1,2}$, 
Pablo Cerd\'a-Dur\'an$^2$, 
Nikolaos Stergioulas$^3$, 
Jos\'e A.~Font$^2$,
\and and Ewald M\"uller$^1$ 
\\
  $^1$Max-Planck-Institut f\"ur Astrophysik,
  Karl-Schwarzschild-Str.~1, D-85741 Garching, Germany \\
$^2$Departamento de Astronom\'{\i}a y Astrof\'{\i}sica,
  Universitat de Val\`encia, 46100 Burjassot (Valencia), Spain \\
$^3$Department of Physics, Aristotle University of Thessaloniki,
  Thessaloniki 54124, Greece
}
\date{\today}
\maketitle

\begin{abstract}
We couple internal torsional, magneto-elastic oscillations of highly
magnetized neutron stars (magnetars) to their magnetospheres.  The
corresponding axisymmetric perturbations of the external magnetic field
configuration evolve as a sequence of linear, force-free equilibria
that are completely determined by the background magnetic field
configuration and by the perturbations of the magnetic field at the
surface. The perturbations are obtained from simulations of
magneto-elastic oscillations in the interior of the magnetar. While
such oscillations can excite travelling Alfv\'en waves in the exterior
of the star only in a very limited region close to the poles, they
still modulate the near magnetosphere by inducing a time-dependent
twist between the foot-points of closed magnetic field lines that exit
the star at a polar angle $\gtrsim 0.19\,$rad. Moreover, we find that
for a dipole-like background magnetic field configuration the magnetic
field modulations in the magnetosphere, driven by internal
oscillations, can only be symmetric with respect to the equator. This
is in agreement with our previous findings, where we interpreted the
observed quasi-periodic oscillations in the X-ray tail of magnetar
bursts as driven by the family of internal magneto-elastic
oscillations with symmetric magnetic field perturbations.
\end{abstract}

\section{Introduction}
Magneto-elastic oscillations of highly magnetized neutron stars
(magnetars) may allow for the first time to infer the interior
properties of these compact objects. Several groups have investigated
their torsional magneto-elastic oscillations (or Alfv\'en oscillations
when the crust is neglected) \citep[see][and references
  therein]{Levin2007, Sotani2008, Cerda2009,
  Colaiuda2009,Colaiuda2011, Colaiuda2011b, vanHoven2011,
  vanHoven2012, Gabler2011letter, Gabler2012, Gabler2013a}. The latest
models even include effects of superfluid neutrons in the core of the
magnetar \citep{Gabler2013b, Passamonti2014}. Most interestingly, the
modulation of the external magnetosphere by these internal
oscillations may have been detected as quasi-periodic oscillations
(QPOs) in the aftermath of two out of the three observed giant flares
of magnetars.  In 2004, the soft gamma-ray repeater SGR 1806-20 showed
the following QPO frequencies in the decaying tail of its giant flare:
$18$, $26$, $30$, $92$, $150$, $625$, and $1840\,$Hz. Oscillation
frequencies have also been found in the 1998 giant flare of SGR
1900+14 at $28$, $53$, $84$, and $155\,$Hz. A strong motivation for
linking internal magneto-elastic oscillations with the observed QPOs
is the fact that several of the observed frequencies appear in a 1:3:5
ratio \citep[as pointed out in][]{Sotani2008}, which cannot be easily
obtained in models that rely on predominantly crustal oscillations,
but which is obtained naturally in the simplest magneto-elastic model
with a purely dipole magnetic field and a regular fluid.

If the magneto-elastic oscillations can explain the observed QPO
frequencies, the open question is: how can these internal oscillations
modulate the emission process in the magnetosphere? A promising
mechanism is the \textit{resonant cyclotron scattering} (RCS) of
photons in the magnetosphere \citep{Timokhin2008}. The fundamental
ingredients for the RCS are the magnetic field configuration, the seed
spectrum of the photons and the scattering targets for photons. In the
magnetosphere the targets are given by the electric currents that are
induced by a twist in the external magnetic field
\citep{Thompson2002}.

In previous work on the RCS \citep{Fernandez2007, Nobili2008, Rea2008,
  Zane2009} the magnetic field was assumed to have a self-similar
solution with possibly multipolar components \citep{Pavan2009}.  More
complicated magnetic field geometries have been studied in
\cite{Vigano2011} and \cite{Parfrey2013}.  \cite{Beloborodov2009}
showed that a twisted magnetic field becomes untwisted by Ohmic
dissipation of the magnetic energy.  In none of these studies the
magnetic field has been obtained consistently with an interior
solution. First equilibrium solutions of coupled interior and exterior
fields with non-vanishing toroidal fields have been recently obtained
in \cite{Glampedakis2014}. However, these equilibria have not been
used for calculations of RCS and, because they are equilibrium
solutions with particular current configurations they cannot be used
for dynamical simulations. To study the interior and exterior magnetic
field evolution of neutron stars in a consistent framework different
groups have developed numerical tools that are based either on
resistive magneto hydrodynamics (MHD) or on some matching between ideal MHD
to its force-free limit \citep{Bucciantini2013, Dionysopoulou2013,
  Palenzuela2013, Paschalidis2013}.

First studies of the coupling of the magnetosphere to internal
oscillations of magnetars have been recently reported in
\cite{Link2014} and \cite{Kojima2014}. The latter use a model of
resistive electrodynamics with artificially low conductivity of the
plasma that is not expected around magnetars \citep{Thompson2002,
  Beloborodov2007, Beloborodov2009, Beloborodov2013long}. An
appropriate description of the magnetosphere of magnetars has to be in
terms of the force-free approximation that we adopt here.

The aim of this paper is to couple the internal magneto-elastic
oscillations of magnetars to their exterior magnetospheres. We
construct linear, force-free magnetic field configurations that are
completely determined by the background configuration and by torsional
perturbations of the magnetic field at the surface, which are obtained
by numerical simulations as in \cite{Cerda2009} and \cite{Gabler2011letter,
  Gabler2012, Gabler2013a, Gabler2013b}.  These configurations are
compared to twisted, self-similar configurations with dipolar
background fields \citep{Thompson2002} to check the validity of our
approach. In our model, the internal oscillations can couple to the
exterior through the closed magnetic field lines and produce
time modulations of the magnetosphere. This is in contrast to
\cite{Link2014}, who finds that the transmission of Alfv\'en waves
from the neutron star crust to the magnetosphere is strongly
suppressed due to the strong impedance mismatch between the two
regions. In this work, we show that this hampered transmission
is only relevant for a small region of open magnetic field lines near
the rotational axis.

The paper is organized as follows: in Section\,\ref{sec_configuration},
we discuss how we construct force-free equilibria magnetic field
configurations in the neutron star exterior by prescribing the axisymmetric
magnetic field perturbation at the surface. The corresponding results
are given in Section\,\ref{sec_results}. In
Section\,\ref{sec_transmission}, we discuss how the interior
oscillations change the exterior magnetic field in the absence of
Alfv\'en waves. Finally, a summary is provided in
Section\,\ref{sec_conclusion}.

\section{Magnetic field configuration}\label{sec_configuration}

The time-scale of magneto-elastic oscillations with frequency
$f\lesssim 150\,$Hz inside a magnetar is of the order of
$t_\mathrm{QPO} \gtrsim 0.007{\rm s}$ \citep{Gabler2012, Gabler2013a,
  Gabler2013b}. In the magnetosphere, the Alfv\'en speed is
approximately equal to the speed of light and thus, out to a distance
of $r \lesssim 1000\,$km, the dynamical time-scale is,
$t_\mathrm{mag}\sim r/c\sim 1/300\,$s~$< t_\mathrm{QPO}$. In the near
magnetosphere, where closed field lines close within a distance of
tens up to a few hundreds of km, $t_\mathrm{mag}\ll t_\mathrm{QPO}$.
Therefore, we assume that the reconfiguration of the external field
occurs on a much shorter dynamical time-scale than the period of the
low-frequency, internal magneto-elastic oscillations (our argument
does not apply to the high-frequency QPOs, which would require
separate considerations). By treating this fast relaxation as if it
effectively occurred instantaneously, the exterior magnetic field
reaches an equilibrium configuration that is determined by the surface
magnetic field.  We thus construct a sequence of static equilibria in
the magnetosphere, and model the modulation of the magnetosphere by
internal oscillations as a quasi-static evolution.  Typical rotation
periods of magnetars $P\sim10\,$s are much longer than the QPO
time-scale and we thus neglect rotational effects.

The magnetic field in the magnetosphere can be assumed to be
force-free. In this approximation, the inertia and momenta of the
charge carriers are neglected with respect to the magnetic field
energy density ($\{\rho,p\}\ll B^2$), where $\rho$ is the density, $p$
is the pressure and $B$ is the magnetic field strength.  Consequently,
the momentum equation leads to the force-free condition
\begin{equation}\label{eq_ff_condition}
 \mathbf{J}\times\mathbf{B} = 0\,,
\end{equation}
where $\mathbf J$ is the current density. This equation states that
the currents have to flow along magnetic field lines, and, hence, that
in equilibrium no Lorentz force is acting on the charge carriers.

Such a configuration can be maintained only under the assumption of
ideal MHD.  In particular, this means that there have to be
sufficiently many charge carriers to make the medium (practically)
perfectly conducting.  In the quiescent state of SGRs, the charge
carriers are provided by a strong and twisted magnetic field. Its
toroidal component $B_\varphi$ creates a large difference in the
electric potential between the foot-points of the field lines that are
anchored in the crust. This potential is sufficiently strong to
accelerate electrons and light ions from the atmosphere just above the
surface of the neutron star (formed by thermally excited particles)
along the magnetic field lines \citep{Thompson2000}. The number of
these charge carriers is by far not sufficient to create the required
currents.  However, particles that are accelerated along the magnetic
field lines create $e^+$-$e^-$ pairs when reaching the energy
threshold for this process. This condition is easily fulfilled in the
case of magnetars. In turn, the pairs get accelerated in the direction
of opposite potential and can create further pairs when having
acquired sufficient kinetic energy. Finally, these pair avalanches
fill the magnetosphere with sufficient plasma to conduct the current
\citep{Beloborodov2007, Beloborodov2013long, Beloborodov2013}. How
this scenario is changed in the case of a giant flare is not clear and
needs further investigation. Here, we assume that the modulation due
to the internal oscillations occurs on field lines that extend to
slightly larger radii $r\gtrsim20\,$km than those field lines that are
expected to host the fireball of the giant flare.

Any static twist of the magnetic field will dissipate on the time-scale 
of years \citep{Beloborodov2007, Beloborodov2009}. This is
orders of magnitude longer than the time-scale of interest for giant
flares ($\sim400 \,$s) and of their QPOs ($\lesssim 4\,$min), i.e. we
can safely neglect dissipation in the magnetosphere, and the
assumption of ideal MHD holds.

\subsection{Self-similar fields}

One solution of equation~(\ref{eq_ff_condition}) is given by currents along
the magnetic field $\mathbf{J} = \mathbf{\nabla} \times \mathbf{B} =
\mathcal{P}(\Gamma) \mathbf{B}$, where $\mathcal{P}$ is a
proportionality factor and $\Gamma$ is a flux parameter
\citep[see][for details]{Thompson2002}.  By making a particular ansatz
for $\Gamma$ one arrives at a \textit{self-similar solution},
i.e. all magnetic field components decay with the same power law
\begin{equation}
 B_i \sim r ^{-2-q}\,,
\end{equation}
where $q$ is an index. The corresponding configuration can be described by the
global twist $\Delta\Phi$ that is defined as the twist angle between the
foot-points of a closed field line that is anchored near the poles
($\theta\rightarrow0$)
\begin{equation} \label{eq_total_twist}
       \Delta \Phi = 2 \int_\theta^{\pi/2} \frac{B_\varphi
(\theta)}{B_\theta(\theta)} \frac{d\theta }{\sin{\theta}}\,.
\end{equation}

In the self-similar model, the choice of $\Gamma$ and of the parameter
$\Delta\Phi$ completely defines the current distribution. However, only
very particular global twisted magnetic fields can be prescribed.  For
small values of $\Delta\Phi \lesssim 0.1$ the configuration remains very
similar to a pure dipole configuration, with $q\sim 1.0$.

\subsection{Force-free magnetic fields in the Schwarzschild spacetime}

The mass of a neutron star causes a significant curvature of the
spacetime and hence also affects the structure of the magnetic field.
For this reason, the internal magneto-elastic oscillations are
computed with a general-relativistic code. To describe the
magnetosphere consistently and to match it to our interior
configurations \citep{Gabler2012, Gabler2013a, Gabler2013b}, we thus
consider a general-relativistic metric using units with $G=c=1$.  In
the exterior, we can assume the metric of a spherically symmetric
star, i.e.\ the Schwarzschild metric
\begin{equation}
 ds^2 = - \alpha^2 dt^2 + \alpha^{-2} dr^2 
                        + r^2 \left(d\theta^2 + \sin^2{\theta}d\varphi^2
                              \right)\,,
\end{equation}
where $\alpha \equiv \left( 1-2M/r \right)^{1/2}$ is the lapse, and
$M$ is the mass of the star. Compared to a Newtonian configuration of
the same mass, the difference in the magnetic field structure near the
stellar surface is of the order of several percent.

We follow \cite{Uzdensky2004} and use an orthonormal basis $e_{k} =
\gamma_{kk}^{-1/2} \partial_k$ with $k = \{r,\theta,\varphi\}$ and
$\gamma_{ij}$ being the 3-metric in the usual $3+1$ split of the
spacetime.  Correspondingly, the 3-dimensional vector operators are
${\nabla} f$ (the gradient), ${\nabla} \cdot \mathbf{B}$ (the
divergence), and ${\nabla}\times\mathbf{B}$ (the curl). We use a tilde
to indicate vector components that are given in the usual co- or
contravariant basis as $ V_{{k}} = \gamma_{kk}^{-1/2} V_{\tilde k} =
\gamma_{kk}^{1/2} V^{\tilde k}$.  The relevant Maxwell equations take
the following form in the $3+1$ split of the Schwarzschild geometry
\citep{MacDonald1982,Uzdensky2004}
\begin{eqnarray}
 {\nabla} \cdot  \mathbf{B}          &=& 0\\
 {\nabla} \times (\alpha \mathbf{B}) &=& \alpha \mathbf{J} \,.
\label{eq_J}
\end{eqnarray}
In this formulation, the Newtonian limit is recovered by setting
$\alpha=1$.

\subsection{Linear reconstruction of $\delta B_{\varphi}$ as 
            a flux function} 
\label{sec_reconstruction}

For an equilibrium background configuration condition
(\ref{eq_ff_condition}) has to be fulfilled. Considering
axisymmetric perturbations up to
linear order, the poloidal components of equation~(\ref{eq_ff_condition})
remain unchanged, while the ${\varphi}$-component has to satisfy the
condition
\begin{eqnarray}
0 &=&\left(\mathbf{J}\times\mathbf{B}\right)_{{\varphi}} 
\nonumber\\
  &=&\frac{1}{r}\left[\frac{B_{{\theta}}}{\sin{\theta}} (\sin{\theta}
      \,\delta B_{\varphi})_{,\theta} + B_{{r}} (r \alpha~\delta
      B_{\varphi})_{,{r}}\right]
\nonumber\\
  &=&\frac{1}{\alpha r \sin{\theta}} \left(\mathbf{B}_0 \cdot {\nabla}
     \right) \left( \alpha r \sin{\theta}\,\delta
     B_{\varphi}\right)\,,
\label{eq_flux_function}
\end{eqnarray}
where $\mathbf{B}_0$ is the poloidal background field, $\delta
B_{\varphi}$ is the magnetic field perturbation in the $\varphi$
direction, and a comma denotes a partial derivative. 
equation\,(\ref{eq_flux_function}) states that $\alpha r
\sin{\theta}\, \delta B_{\varphi}$ does not change in the direction of
the background field, i.e.\ this term is a flux function that is
constant along field lines.

The magnetic vector potential $A_{\tilde\varphi} =
\gamma_{\varphi\varphi}^{1/2} A_{{\varphi}} = r \sin{\theta}
A_{{\varphi}} $ is also a flux function whose equipotential lines
coincide with the field lines. The ${\varphi}$-component of
$0={\mathbf{B}} \times {\mathbf{B}}$ together with the definitions
$B_{ \theta} = -F_{ \varphi r}$, $B_{ r} = F_{ \varphi \theta}$, and
$F_{\tilde\mu\tilde\nu} = A_{\tilde\mu,\tilde\nu} -
A_{\tilde\nu,\tilde\mu}$ give ${\mathbf{B}}\cdot {\nabla}
A_{\tilde\varphi} = 0$, i.e.\ the gradient of $A_{\tilde\varphi}$ is
perpendicular to the magnetic field direction. Hence,
$A_{\tilde\varphi}$ is constant in the direction of ${\mathbf{B}}$,
and we can use $A_{\tilde\varphi}$ to extend $\delta B_{\varphi}$ from
the magnetar surface into the magnetosphere. 

In practice, we consider the value of the potential
$A_{\tilde\varphi}$ at some location $\mathbf{r}_x$ in the
magnetosphere and find the corresponding magnetic field perturbation
$\delta B_{\varphi}$ at the surface location $\mathbf{r}_s$, which has
the same value of $A_{\tilde\varphi}$ by using a 4-point Lagrange
interpolation. In this way, we construct
\begin{equation}
 \delta B (\mathbf{r}) = \frac{\alpha_s r_s \sin\theta_s}{
                               \alpha_x r_x \sin\theta_x} 
                         \, \delta B_{\varphi} (\mathbf{r}_s)\,.
\end{equation}

This approach based on linear reconstruction is very efficient,
because it provides an explicit expression of the
${\varphi}$-component of the magnetic field in the magnetosphere, if
the background poloidal field $\mathbf{B}_0$ or its potential
$A_{\tilde{\varphi}}$ is given.  Moreover, we can directly construct
the field in the magnetosphere from the magnetic field at the magnetar
surface. As the background configuration is known to larger radii than the
numerical domain we consider here, no further boundary conditions
need to be specified at the outer boundary of the numerical grid.

In this work we restrict ourselves to magnetic field configurations
with $\delta B_{\varphi} / B_0 \lesssim 0.1$, for which the linear
approximation is valid. Thus, we can safely neglect corrections of the
poloidal magnetic field by the twist induced by torsional
oscillations.

We obtain the magnetar's surface magnetic field from our simulations
of magneto-elastic oscillations in the magnetar interior
\citep{Gabler2012, Gabler2013a} using background magnetic fields
computed either with the {\small{MAGNETSTAR}} \citep{Gabler2013a}
routine of the {\small{LORENE}}
\footnote{http://www.lorene.obspm.fr} 
library or with an extension of the {\small{RNS}} code
\citep{Stergioulas1995,Friedman2013}. The background model also
provides the poloidal magnetic field in the magnetosphere.

\subsection{Analytic solution in the Newtonian approximation}
\label{sec_ana_magnetosphere}

In a Newtonian framework, we can derive an analytic solution of the
linearized twisted dipole and compare it to the self-similar solution.
For a background dipole (poloidal) field $B_{ r} = 2 m_B \cos{\theta} /r^3$ and
$B_{\theta} = m_B \sin{\theta} / r^3$, where $m_B$ is a measure of the
magnitude of the magnetic field, and by using equation\,(\ref{eq_flux_function}) with
$\alpha=1$, we obtain
\begin{eqnarray}
 0 &=& \frac{1}{r^3} \left[2 m_B \cos{\theta} \left(r~\delta
       B_{\varphi}\right)_{,r} + \frac{m_B}{r} \left(r\sin{\theta}
       ~\delta B_{\varphi}\right)_{,\theta} \right]
\nonumber\\
   &=& 2\cos{\theta} \left(r~\delta B_{\varphi}\right)_{,r} +
       \left(\sin{\theta}~\delta B_{\varphi}\right)_{,\theta}\,.
\end{eqnarray}
 Separating variables as
$\delta B_{\varphi} \equiv f(\theta) g(r)$ gives
\begin{equation}
 2 \frac{[r g(r)]_{,r}}{g(r)} = -
\frac{[\sin{\theta}f(\theta)]_{,\sin{\theta}}}{f(\theta)}\,.
\end{equation}
With the corresponding solution $g(r) \equiv r^{\kappa}$ and $f(\theta) \equiv
\sin^{\lambda}{\theta}$ we obtain the relation
\begin{equation}
 \kappa = -\frac{\lambda + 3}{2}\, ,
\end{equation}
\begin{figure*}[ht!]
\begin{center}  
 \includegraphics[width=.98\textwidth]{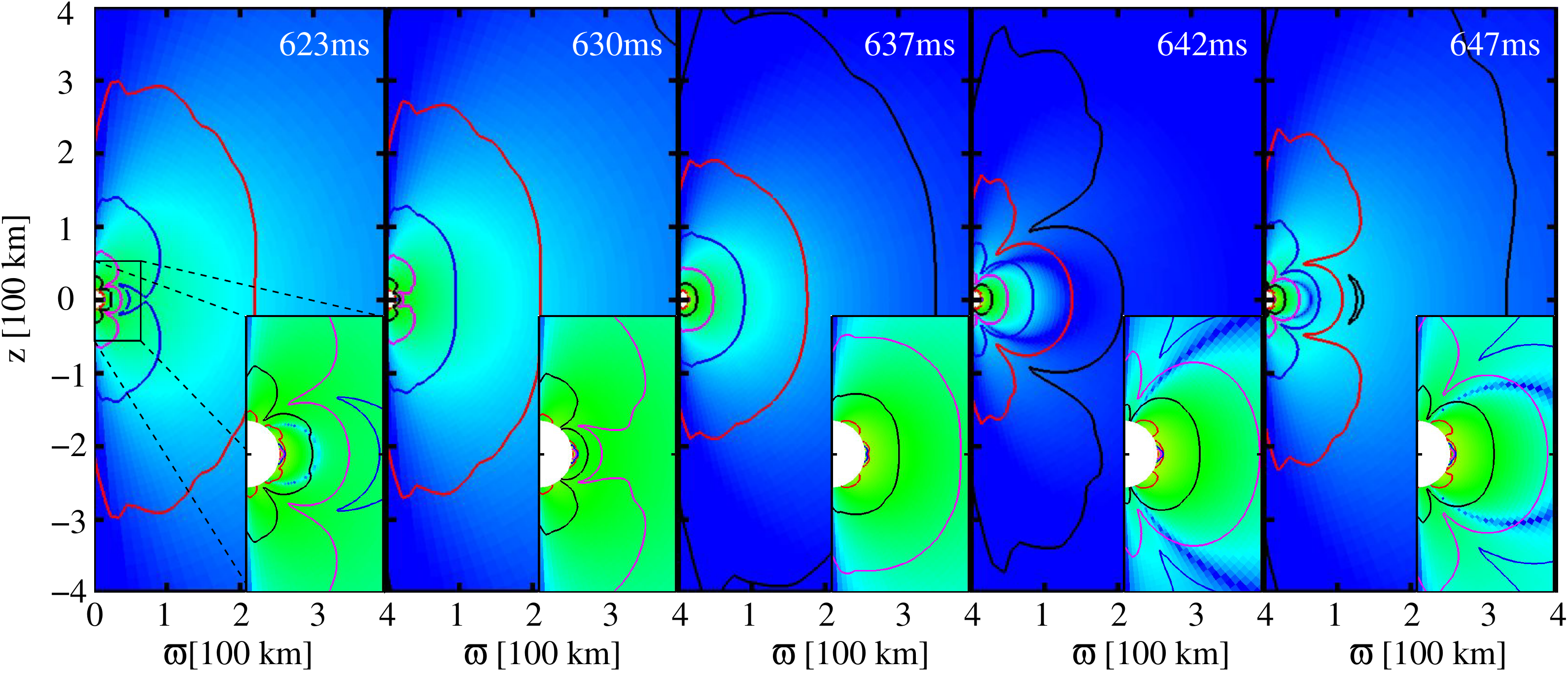}
\end{center}
\caption{Snapshots of the quasi-static evolution of the logarithm of
  the toroidal magnetic field perturbation $\delta B_{\varphi}$ in the
  magnetosphere, matched to an interior simulation of magneto-elastic
  oscillations. The background poloidal magnetic field strength is $B
  = 3\times 10^{15}\,$G, while the toroidal one is $\sim
  10^{14}\,$G. The colour scale is the same in all panels and ranges
  from blue ($10^8\,$G) to orange-red ($10^{14}\,$G). The solid lines
  correspond to constant current surfaces of the absolute value of the
  poloidal current. The snapshot time is given in the top-right corner
  of each panel. The inset in the bottom-right corner of each panel gives a
  magnification of the innermost 50 km.}
 \label{fig_bfield_evo}
 \end{figure*}
and thus
\begin{equation}\label{eq_analytical_sol}
 \delta B_\varphi = r^{-\frac{\lambda+3}{2}} \sin^\lambda{\theta}\, . 
\end{equation}
One sees that $\delta B_{\varphi}$ has a different fall-off behaviour
with $r$ than $B_{ r},~B_{\theta}$, in contrast to the self-similar
solutions that correspond to the particular choice $\lambda = 3$.

\section{Results}
\label{sec_results}

In this section, we discuss the first implications of our model, that
for dipole-like background magnetic fields there can be no
antisymmetric perturbations $\delta B_{\varphi}$ in the
magnetosphere. We then construct configurations that are matched to
simulations of the interior in order to describe the evolution of the
magnetic field consistently. Finally, the range of validity of the
magnetic field reconstruction method is studied.

\subsection{Exclusion of antisymmetric perturbations $\delta B_{\varphi}$
            for dipole-like background magnetic fields}
\label{sec_symmetry_constraint}

In equation~(\ref{eq_flux_function}) we found that $\alpha r \sin\theta
\delta B_{\varphi}$ is a flux function, i.e. this expression has to be
constant along field lines. For dipole-like fields that connect the
two hemispheres of the star and because $\alpha$, $r$ and $\sin\theta$
are all symmetric with respect to the equator, only symmetric
configurations of the torsional magnetic field $\delta B_{\varphi}$
can lead to force-free equilibria.  This holds not only for
dipole-like background fields, but also for more complicated
configurations with field lines exiting and entering the star at
symmetric locations with respect to the equatorial plane.

An antisymmetric perturbation $\delta B_{\varphi}$ implies a symmetric
torsional velocity perturbation $\delta v_{\varphi}$. In this case,
both foot-points of a magnetic field line move with the same speed in
the same direction, i.e.\ the magnetic field line does not become
twisted. In a dynamic (i.e. non-perturbative) calculation the field
line would bend, because $\delta v_{\varphi}$ would change along the
field line. However, there exists no equilibrium solution in this
general situation other than the velocity being zero, and hence
$\delta B_{\varphi} = 0$. This in turn implies that there are no
persistent currents (for $t>t_\mathrm{mag}$) along the field lines and
the emitted radiation will not be modulated.  In contrast, for
symmetric magnetic field perturbations, the twist of the magnetic
field lines is maintained for $t>t_\mathrm{mag}$ because the locations
of their foot-points evolve on much longer time-scales
($t_\mathrm{QPO}>t_\mathrm{mag}$) that are defined by the magnetar's
interior oscillations.

\subsection{Configurations matched to the interior}

The magnetic field configurations in the magnetosphere are obtained
from simulations of the magneto-elastic oscillations of magnetars
\citep{Gabler2011letter, Gabler2012, Gabler2013a, Gabler2013b}. From
these simulations, we obtain (as a function of time) the magnetic
field perturbation $\delta B_{\varphi}$ at the surface of the magnetar
for a given background poloidal magnetic field $\mathbf{B}_0$ in the
magnetosphere.  With this information, we reconstruct the magnetic
field perturbation $\delta B_{\varphi}$ of a force-free equilibrium in
the magnetosphere from equation~(\ref{eq_flux_function}). For our study, we
use an equilibrium model with $R=12.26\,$km, $M=1.4$M$_\odot$ and a
magnetic field strength at the pole of $3\times 10^{15}\,$G. The
equation of state (EOS) is APR \citep{Akmal1998} in the core, matched
to the \cite{Douchin2001} EOS in the crust.

Fig.\,\ref{fig_bfield_evo} displays snapshots from a typical
quasi-static evolution of (the logarithm of) the absolute value of
$\delta B_{\varphi}$ in the magnetosphere towards the end of a
simulation covering $t\sim 650\,$ms. The solid lines are constant current
surfaces of (the absolute value of) the poloidal current consistent
with the toroidal magnetic field. In the magnetosphere, the numerical
grid of $100 \times 80$ ($r \times \theta$) zones covers the range
$[r_\mathrm{s}, 1200\,$km$] \times [0,\pi]$. The radial grid spacing
increases logarithmically, while the angular grid is equidistant. In
Fig.\,\ref{fig_bfield_evo} we show $\delta B_{\varphi}$ up to a
distance $\varpi=400\,$km from the magnetic field axis and up to the same
distance along that axis in positive and negative direction.

At large distances, the magnetic field and the currents in our model
decrease very smoothly, the latter being qualitatively similar to
those of the self-similar solutions (see Section \ref{range}). The
main differences are the stronger decrease of $\delta B_{\varphi}$
with $r$ in the self-similar case ($\lambda \sim 3$), and that for $r
\lesssim 200\,$km the magnetic field can be quite different from that
of the self-similar solutions. As can be inferred from the nodal lines
in last two panels in Fig.\,\ref{fig_bfield_evo}, the magnetic field can change
its sign with increasing radius or with increasing polar angle. This angular
dependence differs from that of the self-similar models, which possess
no nodes in the $\theta$-direction.

\begin{figure}
\begin{center} 
\includegraphics[width=.47\textwidth]{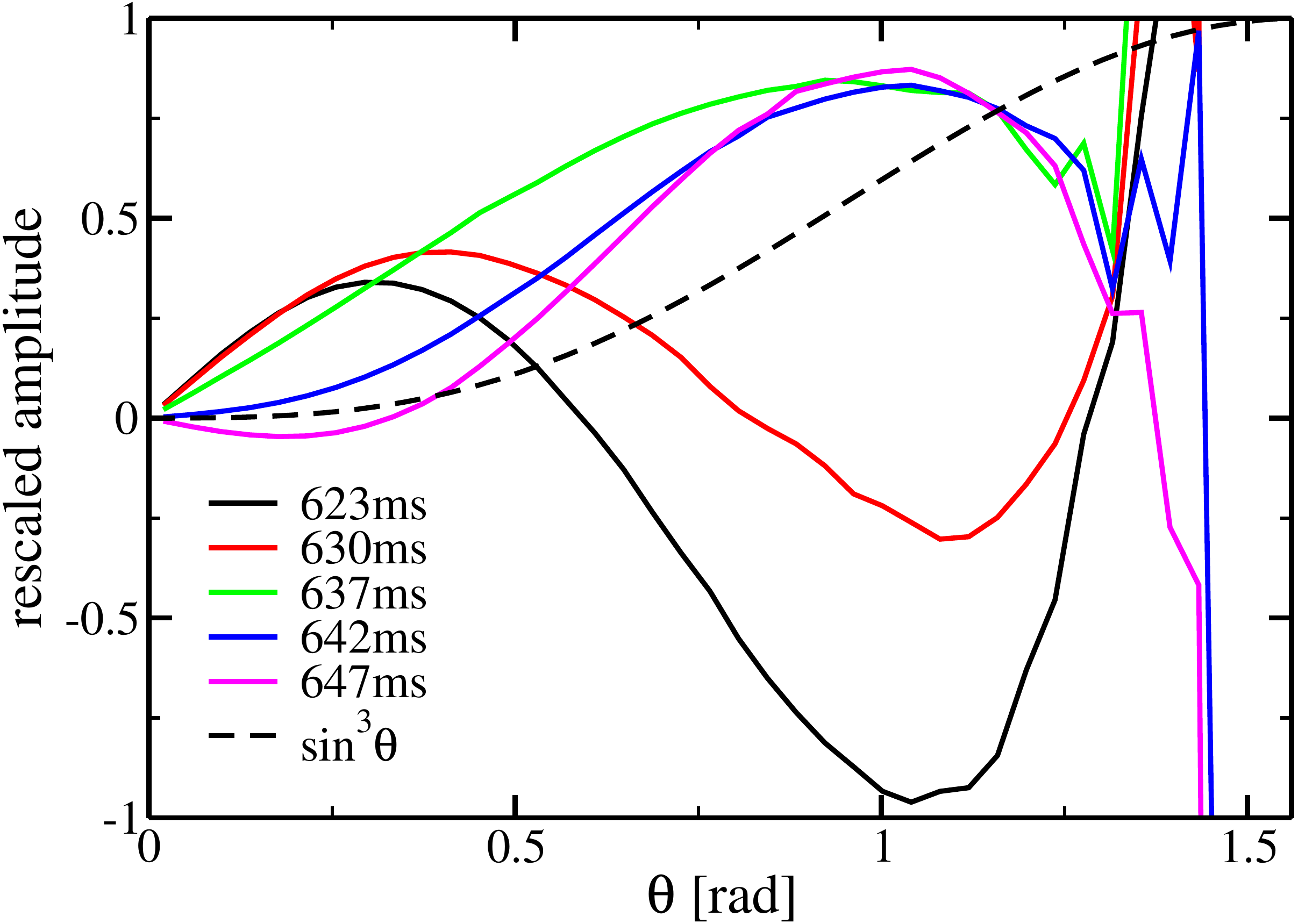}
\includegraphics[width=.47\textwidth]{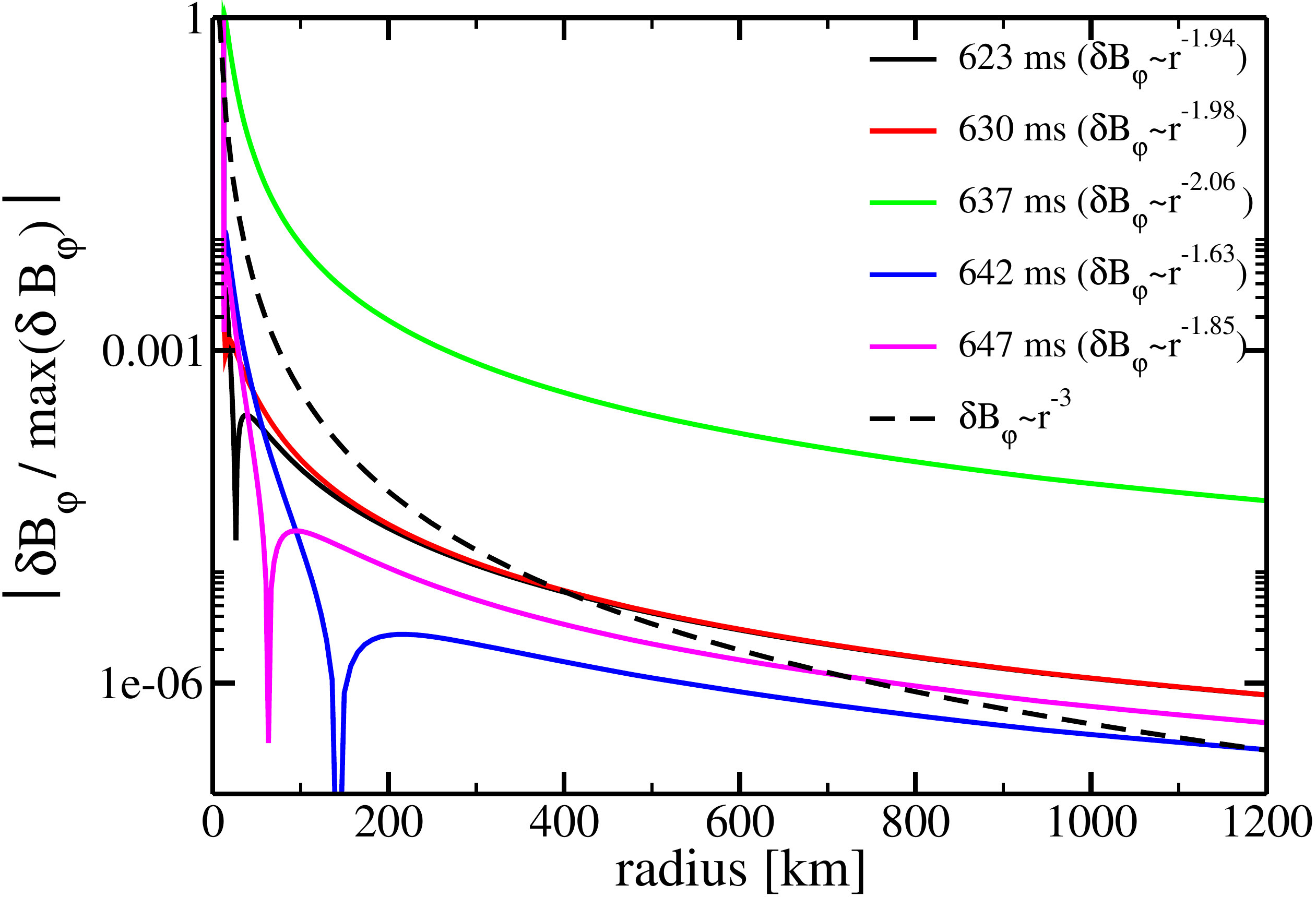}
\end{center}
\caption{{\it Top panel}: snapshots of the magnetic field at the
  surface of the magnetar. The black dashed line corresponds to a
  self-similar solution with angular dependence $\sim \sin^3{\theta}$.
  {\it Bottom panel}: Fall-off behaviour of $\delta B_{\varphi}$ near
  the equator as a function of radius $r$ at different times.  The
  dashed line gives a self-similar solution with $B \sim r^{-3}$ for
  all components of the magnetic field. The asymptotic fall-off of $B$
  for large $r$ is given in the upper right corner of the bottom
  panel.}
 \label{fig_B_r_evo}
\end{figure}

To investigate the magnetic field behaviour near the star, we show the
angular dependence of the rescaled magnetic field at the surface in
the top panel of Fig.\,\ref{fig_B_r_evo} at the same times
used in Fig.\,\ref{fig_bfield_evo}.  The black dashed line in this
panel is an example of a self-similar field with $\delta B_{\varphi}
\sim \sin^3{\theta}$. The figure shows that the magnetic field
structure changes considerably with time.  At $t=623\,$ms, $\delta
B_{\varphi}$ has two nodes (not counting the one at the pole), at
$t=637\,$ms there are no nodes, while at $t=647\,$ms one recognizes
again two nodes.  In contrast, the self-similar field has no nodes.
All configurations show the strongest magnetic field perturbation at
the equator ($\theta = \pi/2$), because only symmetric perturbations
$\delta B_{\varphi}$ are allowed (see Section\,3.1). Therefore, the
velocity $\delta v_{\varphi}$ and hence the displacement
$\xi_{\varphi}$ have to be antisymmetric, i.e.\ they both must have a
node at the equator. This in turn implies that the $\theta$-derivative
of the displacement $\xi_{\varphi,\theta}$ has a maximum at the
equator. The radial derivative of the displacement $\xi_{\varphi,r}$
has to be zero, because for the interior simulations we impose the continuous
traction condition at the magnetar's surface \citep{Cerda2009, Gabler2012}. From
the linearized induction equation 
\begin{equation}
 \delta B_{\varphi} =  B_{ r} \xi_{\varphi\,, r} + B_{\theta} \xi_{\varphi\,,\theta},
\end{equation}
one then finds that $\delta B_{\varphi}$ has a maximum at the
equator. In addition, $B_{\theta}$ has a maximum at the equator too,
which gives rise to a large value of $\delta B_{\varphi}$ close to the
equator. The latter holds, however, only in a small region very close
to the star, because the magnetic field lines of the poloidal
background field originating from the region close to the equator
extend only to about $\lesssim 1\,$km above the surface. The remaining
part of the magnetosphere remains unaffected.

The radial dependence of the magnetic field perturbation $\delta
B_{\varphi}$ at the equator is shown in the bottom panel of
Fig.\,\ref{fig_B_r_evo} at the same times as in
Fig.\,\ref{fig_bfield_evo}.  The dashed black line gives a
self-similar solution that decreases as $\sim r^{-3}$, while the solid
lines are solutions obtained at different times from a given
magneto-elastic simulation.  At small radii, $\delta B_{\varphi}$
decreases for most configurations more rapidly with increasing radius
than the self-similar solution. Only for the configuration at $t =
637\,$ms the decrease is less rapid.  For large radii $r \gtrsim
200\,$km, the magnetic field perturbation $\delta B_{\varphi}$
decreases in all our configurations much slower than the self-similar
solution. The corresponding radial dependences are given in the upper
right corner of the bottom panel of Fig.\,\ref{fig_B_r_evo}. They all
differ significantly from $r^{-3}$.

The radial behaviour of $\delta B_{\varphi}$ at the equator ($\theta =
\pi/2$) reflects its angular behaviour at the magnetar's surface,
because both are linked via the poloidal field lines.  The expression
$\alpha r \sin\theta \delta B_{\varphi}$ has to be constant on these
lines, i.e.\ the strong decrease at small $r$ (see bottom panel of
Fig.\,\ref{fig_B_r_evo}) is related to the strong decrease of $\delta
B_{\varphi}$ at large $\theta$ (see top panel). The nodes of the lines
for $\theta<1.0\,$rad in the top panel correspond to the nodes for
$r<200\,$km in the bottom one. The corresponding values of $r$ and
$\theta$, where the magnetic field possesses nodes, are given in
Table\,\ref{tab_nodes}.  Besides the strong decrease at large angles
all fields, but the one at $637\,$ms (green line), show a node at
about $\theta \gtrsim 1.3\,$rad (see top panel of
Fig.\,\ref{fig_B_r_evo}). The corresponding node in radial direction
is located only a few kilometers above the surface, i.e.\ it cannot be
recognized in the bottom panel.

\begin{table}
\begin{tabular}{c c c c c c c}
Time (ms)& 623&630&637&642&647\\ \hline 
$r$ (km)& 25 & 15& -& 150& 65\\
$\theta$ (rad)&0.6 & 0.8&-&0.05&0.35\\
\hline
\end{tabular}
\caption{Nodes of the magnetic field configurations in the equatorial
  plane (second row) and along the magnetar's surface for
  $\theta<1.0\,$rad (third row) at different times. }
\label{tab_nodes}
\end{table}

The fall-off behaviour of $\delta B_{\varphi}$ at large radii is
determined by its decrease on the magnetar's surface very close to the
pole. The steeper the gradient of $\delta B_{\varphi}$ is as a function
of $\theta$ at the surface, the stronger is the decrease at large
$r$. The steepest fall-offs are obtained at $t=637$, $630$, and
$623\,$ms, respectively (see bottom panel of Fig.\,\ref{fig_B_r_evo}).

\begin{figure*}
\begin{center}
 \includegraphics[width=.9\textwidth]{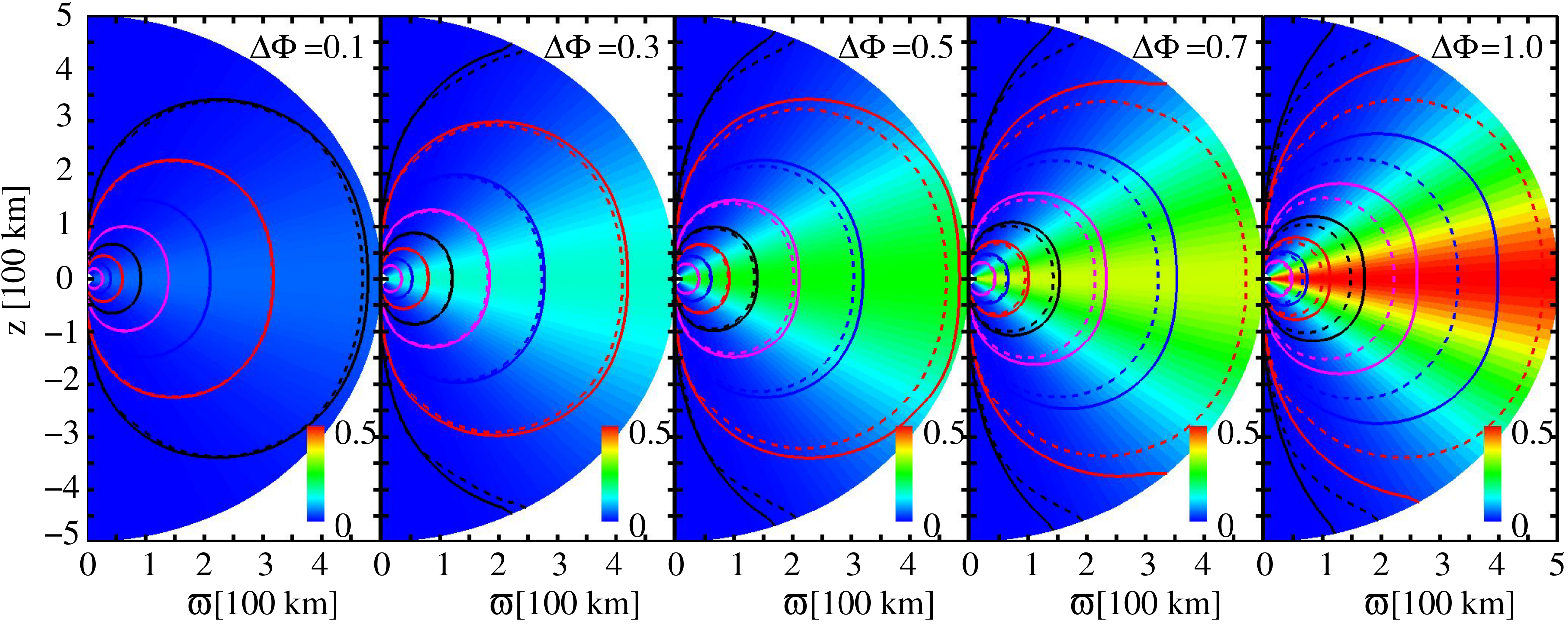}
\end{center}
\caption{Ratio of $\delta B_{\varphi} / B_0$ in the magnetosphere for
  \textit{self-similar} magnetic field configurations (colour coded),
  the {solid lines} showing constant current surfaces. In
  comparison, the {dashed lines} show constant current
  surfaces obtained with the \textit{linear
    reconstruction method}, where the magnetic field at the magnetar's
  surface was chosen to agree with that of the self-similar solution.
  The value for $\Delta\Phi$ given in the different panels denotes the
  total twist and hence a measure of the strength of the toroidal
  magnetic field component relative to the background poloidal
  component (see equation\,\ref{eq_total_twist}).}
\label{fig_ss_compare}
\end{figure*}

\subsection{Range of validity of linear models}
\label{range}

To construct our models, we assume a linear perturbation in $\delta
B_{\varphi}$. In contrast the self-similar solutions are nonlinear,
i.e.\ they can be used to check the validity of our approximation. In
Fig.\,\ref{fig_ss_compare} we show the ratio of toroidal and poloidal
magnetic field strength, $\delta B_{ \varphi}/B_0$, for different
self-similar solutions with twist angles $\Delta
\Phi=\{0.1,0.3,0.5,0.7,1.0\}$. With increasing $\Delta \Phi$, the
relative strength of the toroidal field component increases and
reaches the same order of magnitude as the poloidal one at the
equator, for $\Delta\Phi = 1.0$ . In a nonlinear treatment, such a
strong toroidal magnetic field would lead to an inflation of the
poloidal component \citep{Roumeliotis1994, Vigano2011}.

The solid lines in Fig.\,\ref{fig_ss_compare} represent constant poloidal
current surfaces that is consistent with the toroidal
magnetic field component of a self-similar solution, while the dashed
lines are obtained with our linear reconstruction method. The same
colours representing the same current magnitudes. For a weak twist
($\Delta\Phi = 0.1$, first panel) the lines for the self-similar and
linear method are almost indistinguishable, i.e. both methods give
approximately the same currents.  The stronger the twist, the more the
linear results differ from the self-similar ones. For the strongest
twist shown in the last panel ($\Delta\Phi = 1.0$) the constant poloidal current
surfaces differ significantly among the two
approaches. In the linear case we underestimate the currents, because
we neglect the toroidal currents that are present in the nonlinear,
self-similar configurations. Additionally, the poloidal magnetic field
inflates in the latter case, i.e.\ the poloidal currents extend
further into the magnetosphere. Therefore, the blue dashed line
(linear method) crosses the equator closer to the star at
$\sim330\,$km than the solid line (self-similar solution) that crosses
at $\sim400\,$km. Despite these quantitative differences, the general
shape of the constant current surfaces is very similar for both the
linear approximation and the self-similar solution.

Based on the differences in the currents, the configuration with
$\Delta\Phi=0.5$ can be regarded as the limiting case up to which one
can apply the linear reconstruction method.  For this configuration
the ratio of the toroidal and poloidal magnetic field strength is less
than $25\%$ and the current amplitudes at $r=100\,$km of the
reconstructed field and the self-similar solution differ by less than
$5\%$. In our models we employed a \textit{more conservative estimate}
of the maximum acceptable toroidal magnetic field and assumed the
linear approximation to be valid up to $\delta B_\varphi / B_0 \leq
0.1$.

\begin{table}
\begin{tabular}{c c c c c c}
Time (ms)& 623&630&637&642&647\\ \hline 
${\delta B_{\varphi}}/{ 
B_0} $&0.08&0.06&0.02&0.03&0.06\\
\hline
\end{tabular}
\caption{Maximum values of $\delta B_{\varphi} / B_0$ for the five
  quasi-stationary configurations considered in our study. The maxima
  are always located at the outer boundary of the computational
  domain at $r \sim 1200$ km and $\theta = \pi/2$.}
\label{tab_referencemodel}
\end{table}

We give the maximum of $\delta B_\varphi / B_0$ for the five magnetic
field configurations displayed in Fig.\,\ref{fig_bfield_evo} in
Table\,\ref{tab_referencemodel}. It never exceeds a value of $0.1$ in
the computational domain that extends up to $r=1200\,$km, and is
always located at the most distant points of the computational grid
in the equatorial plane. This is a consequence of the fact that the
fall-off of the toroidal magnetic field perturbation $\delta
B_{\varphi} \sim r^{-1.5 \dots-2.1}$ is less steep than that of the
poloidal background field $B \sim r^{-3}$. The linear reconstruction
method is thus a very good approximation up to distances of $r\sim
1000\,$km.

\section{Transmission of Alfv\'en waves}
\label{sec_transmission}

\begin{figure*}
\begin{center}
 \includegraphics[width=.9\textwidth]{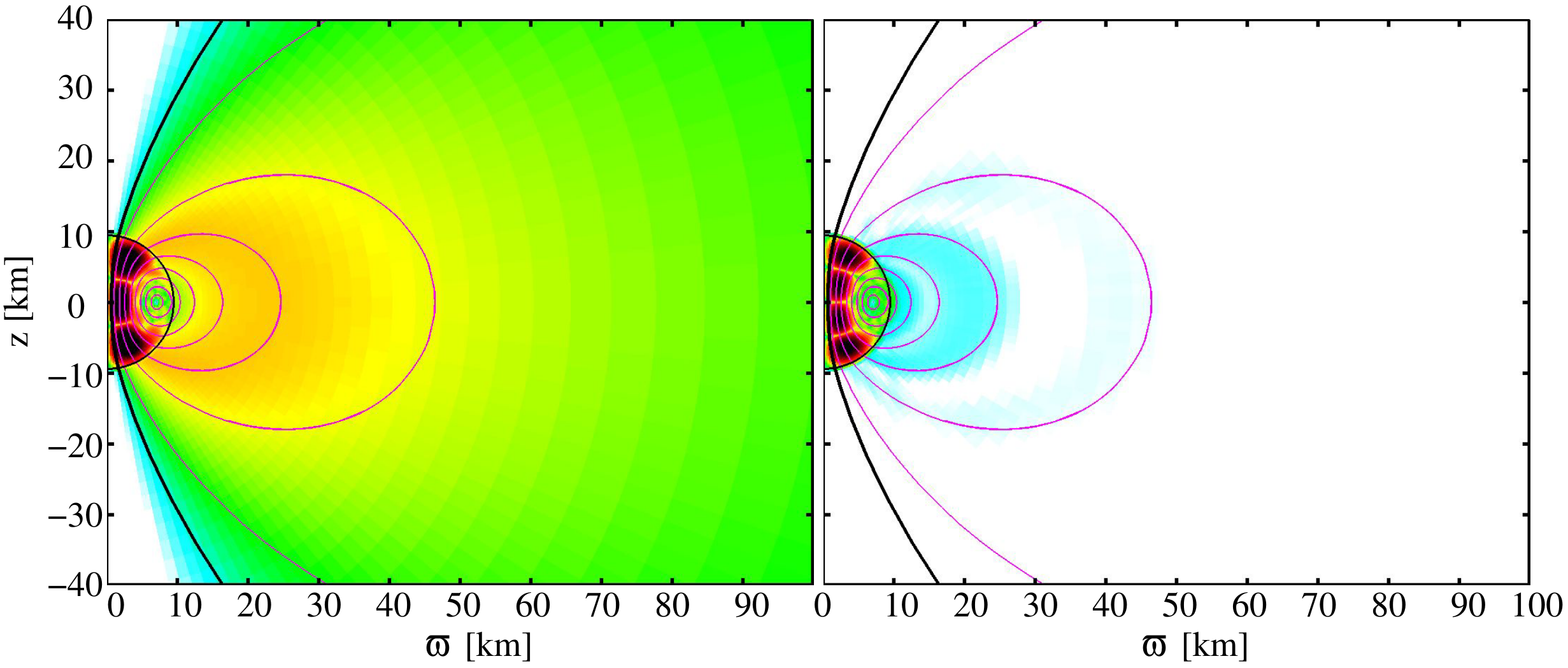}
\end{center}
\caption{Rescaled absolute value of the Fourier amplitude of the
  evolution of the toroidal magnetic field perturbation $\delta
  B^{\tilde\varphi}$ for a perturbation with a frequency of $f=22\,$Hz
  that is symmetric (left-hand panel) and with a frequency of
  $f=29\,$Hz that is antisymmetric (right-hand panel) with respect to
  the equator. The solid lines show magnetic field lines. To the right of
  the solid thick black line is the region where Alfv\'en waves with
  $f < 150\,$Hz cannot propagate.  The logarithmic colour scale ranges
  from white-turquoise ($10^{-7}$), green ($10^{-4}$), yellow
  ($10^{-3}$) to red-black ($1$). The perturbation of the exterior
  magnetic field is orders of magnitude smaller in case of an
  antisymmetric perturbation inside the star (right-hand panel) than in
  case of a symmetric one (left-hand panel). }
\label{fig_QPO_ext}
\end{figure*}

In a recent paper, \cite{Link2014} studied the transmission of
torsional Alfv\'en waves from the magnetar's interior into its
magnetosphere. He found that most of the energy of the oscillations
cannot be transmitted, because of the very different propagation
speeds in the two regions. However, this conclusion only holds in his
plane-parallel toy model, where waves can be excited in the exterior
along open magnetic field lines. As we show next, \textit{when a more
  realistic model of a global dipolar magnetic field is considered,
  the considerations of \cite{Link2014} are relevant only for field
  lines very close to the magnetic poles}.

To allow for torsional Alfv\'en wave transmission along magnetic field
lines, the lengths of the field lines $l_\mathrm{fl}$ have to be
larger than the wavelength $\lambda_f$ of an oscillation at given
frequency $f$.  At $f \sim 150\,$Hz the corresponding wavelength is
$\lambda_f \sim \frac{c}{f} \gtrsim 2000\,$km. Assuming a
dipole-like background magnetic field configuration, the corresponding
limiting field line has a maximum extent of approximately $\varpi
\sim300\,$km, and intersects the magnetar's surface approximately at
$\theta \sim 0.19\,$rad.  Thus, Alfv\'en wave transmission can take
place only at very small polar angles, i.e.\ in a narrow cone along
the axis of the magnetic field.  Our estimate $\theta \lesssim
0.19\,$rad is rather conservative, because even for $l_\mathrm{fl} >
\lambda_f$ the waves cannot travel freely along magnetic field
lines. Closed magnetic field lines that are anchored to the crust can
only be excited to oscillate at certain frequencies. The latter depend
on the length of the field line, the magnetic field strength, and the
boundary condition imposed at the stellar surface. For a complete
description one would have to solve the coupled
core-crust-magnetosphere problem which is computationally expensive.
However, for the frequencies of interest here, $f \lesssim 150\,$Hz,
and thus for field lines exiting the star at $\theta \gtrsim
0.19\,$rad, we can apply the approximations described above.

We confirmed the correctness of our argumentation by performing a
simulation (extending up to $t=1\,$s) of internal Alfv\'en
oscillations with the GRMHD code {\small MCOCOA} \citep{Cerda2008,
  Cerda2009, Gabler2012}. The simulation setup includes a very low
density, artificial atmosphere which extends up to $r \lesssim
1000\,$km.  The density at the surface is set to $\rho_\mathrm{s} =
10^{-10} \rho_\mathrm{center} \sim 4\times 10^5\,$g cm$^{-3}$. To
guarantee an approximately constant Alfv\'en velocity $v_\mathrm{A}
\sim c$ in the atmosphere, the density falls off as $r^{-4}$. For the
simulation, we use our fiducial equilibrium model with a dipole field
strength of $B=10^{15}\,$G and neglect the crust. In the outer parts
of the crust, magnetic forces are much stronger than shear forces,
i.e.\ the magnetic field dominates the coupling to the exterior.
Therefore, the influence of the crust can be safely neglected even
though its presence could affect the structure of particular internal
oscillations and could also shift their frequencies by some small
amount.  The initial perturbation of the equilibrium configuration has
a mixed $l=2$ and $l=3$ angular dependence. At the surface of the
magnetar we need not to prescribe boundary conditions, because the
continuous traction condition used by \cite{Cerda2009} and \cite{Gabler2012} is
equivalent to momentum conservation which is guaranteed by our
numerical MHD scheme. At the outer boundary, we explicitly apply the
continuous traction condition which does not allow for surface currents
\citep[see][for details]{Cerda2009, Gabler2012}.

Fig.\,\ref{fig_QPO_ext} shows the (absolute value of the) Fourier
amplitude of the toroidal magnetic field component $\delta
B^{\tilde\varphi}$ that is created in the magnetosphere by the
coupling to an internal Alfv\'en oscillation.  The left-hand panel shows
the amplitude corresponding to a particular oscillation with a
frequency of 22 Hz that is symmetric (in $\delta B^{\tilde\varphi}$)
with respect to the equatorial plane, while the right-hand panel displays
an antisymmetric oscillation at $f=29\,$Hz. Note that in our previous
work \citep[e.g.][]{Cerda2009, Gabler2011letter, Gabler2013a} we used
the velocity to compute the Fourier amplitude, and that the velocity
has always a maximum where the magnetic field has a node and vice
versa.  Travelling Alfv\'en waves with $f \lesssim 150\,$Hz can be
transmitted only in the region to the left of the thick black magnetic
field line in the figure, where the wavelength
$\lambda_{150\mathrm{\,Hz}} < l_\mathrm{fl}$.

In the near magnetosphere ($r \lesssim 100\,$km), the external field
relaxes almost instantaneously (compared to the interior evolution
time-scale) to a force-free equilibrium, whose configuration is
determined by the structure and the amplitude of the perturbation at
the magnetar's surface (see previous sections). The resulting magnetic
field structure does not resemble one that could be produced by
standing Alfv\'en waves. The latter would give rise to nodes along the
field line, whereas each configuration considered here has a static
twist $\delta B^{\tilde\varphi} \neq 0$ everywhere along a given field line,
i.e. there are no nodes. From
equation\,(\ref{eq_flux_function}), we expect $\alpha \delta
B^{\tilde\varphi} = \alpha r \sin\theta \delta B_{\varphi}$ to be
constant along magnetic field lines. Although the interior
oscillations of the magnetar shown in the two panels of
Fig.\,\ref{fig_QPO_ext} have different symmetries, we find that the
condition $\delta B^{\tilde\varphi} = \mathrm{const.}$ holds
approximately in both cases. Small deviations at small radii are
caused by the factor $\alpha$ whose influence we do not consider in
this figure.

Fig.\,\ref{fig_QPO_ext} also shows that for antisymmetric interior
oscillations (right-hand panel) the amplitude of the exterior magnetic
field perturbation is orders of magnitudes smaller (white and
turquoise regions) than for symmetric oscillations (left-hand panel).  In
addition, we note that outside the star $\delta B^{\tilde\varphi}$ is
symmetric with respect to the equator, too
\footnote{This is an artefact of the FFT which has a limited
  resolution in the frequency domain, resulting in an overlap of
  oscillations at frequencies very close to a given frequency.}.
Our simulations confirm the result of our linear reconstruction method
that antisymmetric oscillations cannot be communicated to the exterior
in the near magnetosphere.

\begin{figure}
\begin{center}
 \includegraphics[width=.47\textwidth]{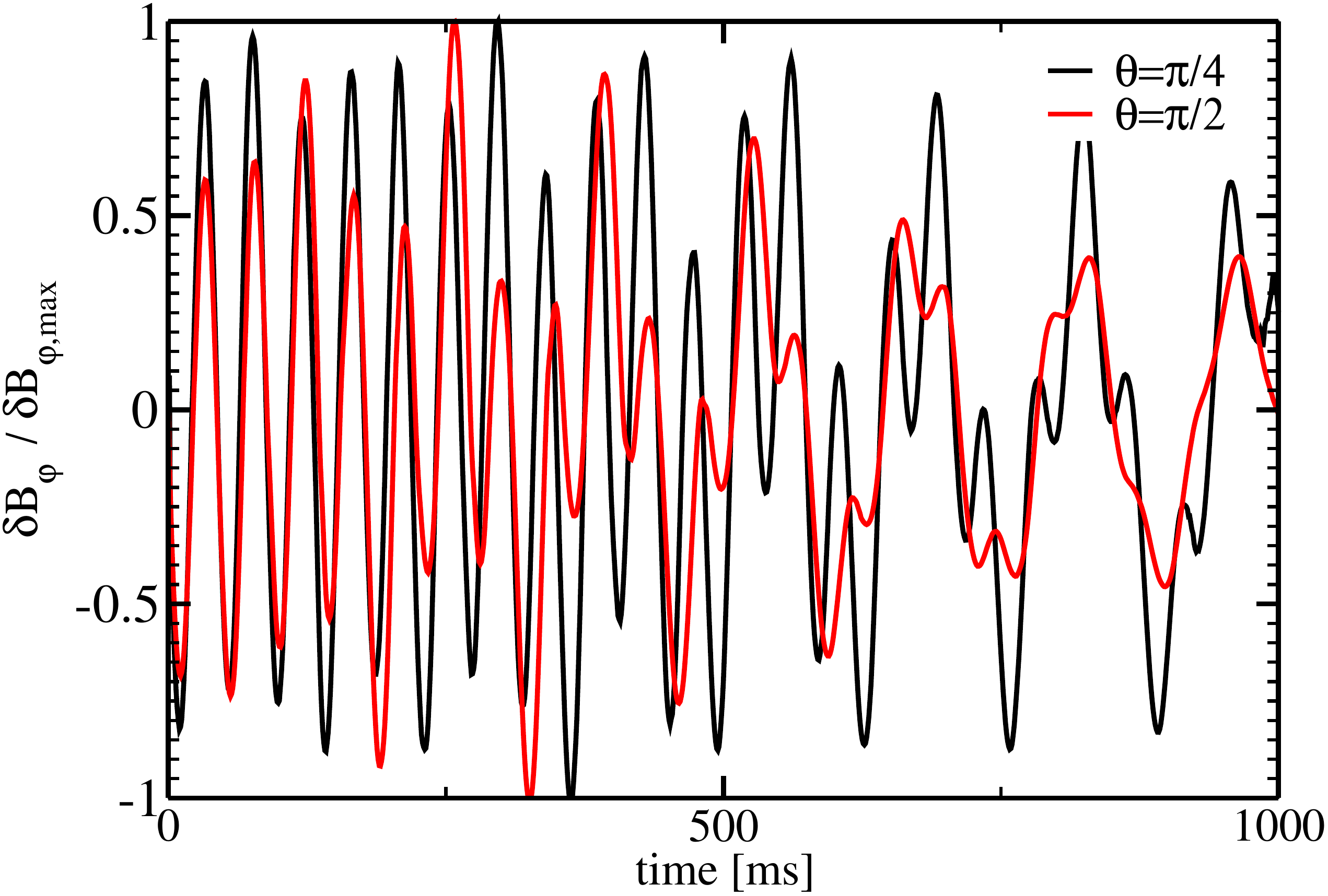}
\end{center}
\caption{Time evolution of $\delta B_\varphi$ at $r=51\,$km and two
  polar angles $\theta$ rescaled to its maximum value.  }
\label{fig_bf_51km}
\end{figure}

In Fig.\,\ref{fig_bf_51km}, we illustrate the evolution of the magnetic
field perturbation $\delta B_\varphi$ at $r=51\,$km and $\theta =
\pi/2$ and $\theta=\pi/4$, respectively. The exterior field is clearly
modulated at the frequencies that are determined by the interior
oscillations.

The study presented in this section shows that the amplitude of the
modulations of the near magnetosphere caused by internal oscillations
is not limited by the reflection of plane-parallel waves as calculated
by \cite{Link2014}. Instead, the amplitude of the modulations is
determined by how much the exterior field is twisted by the motion of
the foot-points of the magnetic field lines that are anchored to the
crust. These quasi-static modulations of the magnetospheric field can
lead to modulations in the X-ray emission and thus may cause the
observed QPOs, through e.g.\ the resonant cyclotron scattering
process.

\section{Conclusions}\label{sec_conclusion}

In this paper we have investigated how the low frequency ($f \lesssim
150\,$Hz) torsional magneto-elastic oscillations of a magnetar can
modulate the exterior magnetic field. In addition, we have shown that
these oscillations are relevant for Alfv\'en wave transmission only
along field lines that arise within a very narrow cone ($\theta
\lesssim 0.19\,$rad) around the polar axis. Here, we are mainly
interested in what happens during a giant flare that is supposed to
produce a fireball close to the magnetar's surface
\citep{Thompson2001}.  In this region, the magnetospheric field finds
its force-free equilibrium configuration much faster than the internal
oscillation time-scale, i.e.\ the magnetosphere evolves
quasi-statically through a sequence of force-free equilibria.
However, at any given time the internal oscillations determine the
magnetic field at the magnetar's surface. The shift of the foot-points
of the external magnetic field lines relative to the unperturbed
configuration twists the external magnetic field, i.e.\ the field is
no longer a potential field and currents flow in the magnetosphere.

For a dipole background magnetic field we have shown that only axisymmetric,
torsional magnetic field perturbations that are symmetric with respect
to the equator are allowed (the corresponding velocity perturbation is
antisymmetric). That only symmetric perturbations are viable is
a promising result, because the frequencies of the low-frequency
QPOs observed in the two giant flare sources come in a near $1:3:5$
ratio, as pointed out first by \cite{Sotani2008} who, however, studied only
purely Alfv\'en oscillations.  For the case of global magneto-elastic
oscillations we found \citep{Cerda2009, Gabler2012} that for
sufficiently strong dipole magnetic fields (when the oscillations can
penetrate the crust and reach the surface) symmetric oscillations of
$\delta B_{\varphi}$ also have approximately the same odd-integer
$1:3:5$ frequency ratio. In contrast, symmetric velocity perturbations would
lead to a uni-directional shift of the field lines in the azimuthal
direction.  In this case, no toroidal magnetic field component would
be created, and the only allowed equilibrium solution has a vanishing
$\delta B_{\varphi}$.

We have shown how magneto-elastic oscillations can modulate
quasi-statically the magnetosphere and how the corresponding magnetic
field configurations can be obtained instantaneously from the
perturbations at the magnetar's surface. The next major step towards a
direct connection between the theoretical modeling and the observed
QPOs in magnetar giant flares consists in taking into account an
emission mechanism for which the modulations of the magnetosphere
cause the observed variations of the light curve in the X-ray band. A
promising candidate is resonant cyclotron scattering, as already
pointed out in \cite{GablerThesis} and \cite{Gabler2014}.

\section*{acknowledgements}

Work supported by the Collaborative Research Center on Gravitational
Wave Astronomy of the Deutsche Forschungsgemeinschaft (DFG
SFB/Transregio 7), the Spanish {\it Ministerio de Educaci\'on y
  Ciencia} (AYA 2010-21097-C03-01), the {\it Generalitat Valenciana}
(PROMETEO-2009-103 and PROMETEO-2011-083), and the EU through the ERC
Starting Grant no. 259276-CAMAP and the ERC Advanced Grant
no. 341157-COCO2CASA. Partial support comes from NewCompStar, COST
Action MP1304. Computations were performed at the {\it Servei
  d'Inform\`atica de la Universitat de Val\`encia}.

\bibliographystyle{mn2e_new}
\bibliography{magnetar}

\end{document}